# Street Hierarchies: A Minority of Streets Account for a Majority of Traffic Flow

Bin Jiang

Department of Land Surveying and Geo-informatics
The Hong Kong Polytechnic University, Hung Hom, Kowloon, Hong Kong
Email: bin.jiang@polyu.edu.hk

**Abstract**
Urban streets are hierarchically organized in the sense that a majority of streets are trivial, while a minority of streets is vital. This hierarchy can be simply, but elegantly, characterized by the 80/20 principle, i.e. 80 percent of streets are less connected (below the average), while 20 percent of streets are well connected (above the average); out of the 20 percent, there is 1 percent of streets that are extremely well connected. This paper, using a European city as an example, examined, at a much more detailed level, such street hierarchies from the perspective of geometric and topological properties. Based on an empirical study, we further proved a previous conjecture that a minority of streets accounts for a majority of traffic flow; more accurately, the 20 percent of top streets accommodate 80 percent of traffic flow (20/80), and the 1 percent of top streets account for more than 20 percent of traffic flow (1/20). Our study provides new evidence as to how a city is (self-)organized, contributing to the understanding of cities and their evolution using increasingly available mobility geographic information.

## 1. Introduction

As a basic man-made infrastructure and backbone of cities, urban streets demonstrate a hierarchical structure in the sense that a majority is trivial, while a minority is vital. This hierarchy can be reflected in the city maps; there are always far fewer yellowish streets than gray ones with Google Maps at a city level, i.e., smaller streets are far more common than larger ones. Hierarchy is also one of the principles of cartographic design, i.e., those prominent in reality should look prominent as well in the maps. Hierarchy is illustrated to exist in many natural and social systems such as cells, cities, the Internet, and languages (Pumain 2006). Although streets and patterns have been an intriguing research topic (Marshall 2004), little quantitative evidence is provided as to how urban streets are hierarchically organized. Such quantitative studies of street hierarchies are only made possible nowadays as massive geographic information has been becoming available. In this respect, geographic information collected about cities is remarkably rich, in particular, the increasing availability of mobility geographic information from mobile devices such as cell phones equipped with Global Positioning System (GPS) receivers. The geographic information about cities at very fine levels will drive the study of cities using a complexity paradigm, essentially a bottom-up approach focusing on interaction between constituents.

Streets are not independent entities, but intersected and connected to form a network topology, i.e. street-street topology. Data mining from the street-street topology is of significant value towards the understanding of street hierarchies from a topological perspective. The previous study (Jiang 2007) illustrated the fact that urban streets demonstrate a scaling law (Zipf 1949), and can be characterized by the 80/20 principle, i.e. 80 percent of streets are less connected (below the average), while 20 percent of streets are well connected (above the average); out of the 20 percent, there is 1 percent of streets that are extremely well connected. Tomko and his colleagues (2008) have recently examined how the hierarchies exist and are used in human perception and cognition, i.e., how the prominent streets constitute a common knowledge in human wayfinding and route descriptions. Although it is not a street pattern in a very restricted sense, Carvalho and Penn (2004) also found that maximum visibility lines (a sort of street segments) follow a scaling law, a hierarchy from a geometric perspective. Inspired by the previous studies, this paper aims to examine the hierarchies of urban streets at a much more detailed level, from the perspective of both geometry and topology, and from the point of view of traffic flow in reality. We illustrated some intriguing street hierarchies and compared them against the intensity of traffic flow collected from taxi cabs equipped with GPS receivers. We found to our surprise that the street hierarchies conform pretty well to the intensity of traffic flow in reality. Our study provides new evidence as to how a city is (self-)organized, contributing to the understanding of cities and their evolution using increasingly available mobility geographic information.

The rest of this paper is structured as follows. In section 2, we introduce power law distributions to mathematically characterize the hierarchies of complex systems. Section 3 examines the street hierarchies of Gävle city (a city in east central Sweden, about 1.5 hours by train north of Stockholm) from the perspective of geometric and topological properties. The hierarchy is further investigated in Section 4 from the perspective traffic flow, revealing the fact that a minority of streets account for a majority of traffic. Finally section 5 discusses the implications of our findings and concludes the paper.

**2. Power laws, Zipf's law, and Pareto Distributions**
A quantity follows a power law when the probability or frequency of its value varies inversely as a power of that value. Many natural and man-made phenomena exhibit the power laws in the way that there are many small events, and few large events. For instance, that there are many earthquakes that are no damaging at all, but a few that are extremely damaging, like the earthquake off the coast of Northern Sumatra and its resultant tsunami; that many websites are much less visited, but a few like Google and Yahoo are extremely well visited; that there are a few mega cities, but many small cites or towns. In all these examples, there is a quantity to measure occurrences of the phenomena. Thus that value (*x*) of the quantity and probability (*p(x)*) of that value form a power law:

$$p(x) = kx^{-\alpha} \qquad [1]$$

where both *k* and α are constants.

The histogram of the power law is highly right-skewed, as shown in Figure 1, where the *x* axis is the rank of streets, while the *y* axis is street length. The curve can be approximately represented by $y = 121669x^{-1.02}$. This is taken from the following experiment about street length of the Gävle street network. The power law is often called heavy tail or long tail distribution, indicating the fact that there are many more shorter streets than longer ones. In this respect, the mean or average length of streets makes little sense, because of the diversity of street length. In fact, power laws are much more popular (or normal) than Gaussian (or normal) distributions, and are often called a universal law, a well received study in fields such as physics, economics, and biology. With the increasing availability of massive data about topological and hyperlink structure of the Internet and WWW, power laws have received a revival of research interests in an array of disciplines.

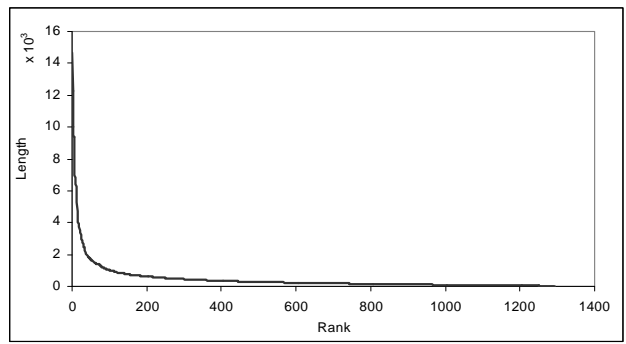

Figure 1: Power law distribution of street length

The observation of power laws was made first by Auerbach (1913) (quoted in Newman 2005), but was popularized first by Zipf (1949) for city size distributions, word frequencies, and income distributions. The frequency of English words occurred unevenly in a text. That is, some words such as "the" and "a" occur very frequently, while the remaining vast majority of words occur extremely rarely. More accurately, Zipf found that the occurrence of an event is in inverse proportion to its rank; that is, proportional to 1, 1/2, 1/3, ¼, etc. For instance, if the most frequent word occurs 100 times in the text, then the second most frequent word would occur 50 times, the third 33 times and so on and so forth. In ranking world cities, if the largest city has a population of 10M, then the second largest city will have 5M, … and the fifth largest 2M, etc. The rank-size distribution, or Zipf's law, can be expressed as the probability distribution function (PDF),

$$p[X = x] \sim x^{-\alpha} \qquad [2]$$

where x is the rank, and $\alpha$ a constant close to 1.

In examining the distribution of income, the Italian economist Vilfredo Pareto was interested in how many people have an income greater than *x*. In contrast to Zipf's law, Pareto distribution is concerned about the cumulative distribution function (CDF).

$$p[X > x] \sim x^{-k} \qquad [3]$$

Note the difference between *X = x* and *X > x* with respect to [2] and [3].

To examine the existence of a power law, $p(x) = kx^{-\alpha}$ in general, the standard way is to plot the logarithms of that value (*x*) and its probability *p(x)*, i.e., $\log(p(x)) = -\alpha \log(x) + k$. The straight line on the log-log plot is the signature of a power law. However, in reality the straight line is hardly observed over the entire range of the value (*x*), and most power laws instead are obtained by a cutoff at the smallest perceivable size (Newman 2005).

There are many terms to refer to the power laws in the literature, such as scaling, scale invariance, scale-free, universality, hierarchy, heterogeneity, and nonlinearity. Research on the origins of power laws, and efforts to observe and validate them in the real world, is particularly active in many fields of science including for instance, physics, economics, biology and more recently computer science. Enormous theories have developed to explain the generative mechanisms of power law distributions. Of particular note are (1) the rich get richer mechanism, firstly developed by Simon (1955), and rediscovered by Barabási and Albert (1999), the so called preferential attachment for the growth of the Internet and web; and (2) self-organized criticality built on the famous sandpile model (Bak, Tang and Wiesenfeld 1987, Bak 1996). The theory of self-organized criticality states that when a system is approaching an unbalanced status, its behavior shows complexity, signified by a power law distribution. Both theories shed light on the formation of streets or cities in general. The existence of power laws is one of the most striking signatures that cities are a product of self-organization (Krugman 1996, Portugali 2000). On the other hand, power laws are often thought to be signatures of hierarchy and robustness. This is what we intend to explore in the paper to do with urban streets – street hierarchies.

**3. Street hierarchies of Gävle city**
We chose the Gävle street network for our study of street hierarchies, because of the availability of the dataset and GPS tracking logs. Based on the network, both named streets (Jiang and Claramunt 2004) and natural streets (Thomson 2003, Jiang and Liu 2008) are derived for further analysis. The natural streets are naturally merged street segments with a good continuity according to the Gestalt principle. The named streets are formed according to identified names. However, there is no guarantee that every street segment has a street name attached to it, because of incompleteness of street network databases. Those segments without names are merged in terms of a good continuity in this study. At an operational level, we compare neighboring segments and merge those segments with the smallest deflection angle. In this particular study, the threshold angle for the merging process is set as 60 degrees, which implies that the deflection angle between two adjacent segments greater than 60 should not be merged for forming a part of a natural street. The named streets are generated simply by dissolving street segments with the same name. For further details about the algorithms for forming natural and named streets, the reader should refer to (Jiang and Liu 2008). This operation applied to the Gävle street network leads to 1292 natural streets and 1002 named streets. The reason the number of natural streets exceeds that of named streets is mainly due to the fact that some major streets are separated by a barrier, forming two or more natural streets. The distribution of street length with respect to its rank is strongly skewed. As shown figure 2, the curves with respect to two types of streets in the Zipf plot have two parts: the tilted part (upper) and the vertical part (lower), divided around the length of 100 meters. It implies that those streets longer than 100 meters demonstrate regularity, whereas those shorter than 100 meters show a different behavior. In other words, the streets longer than 100 meters are strongly differentiated in length. We will elaborate on the issue later on in the paper.

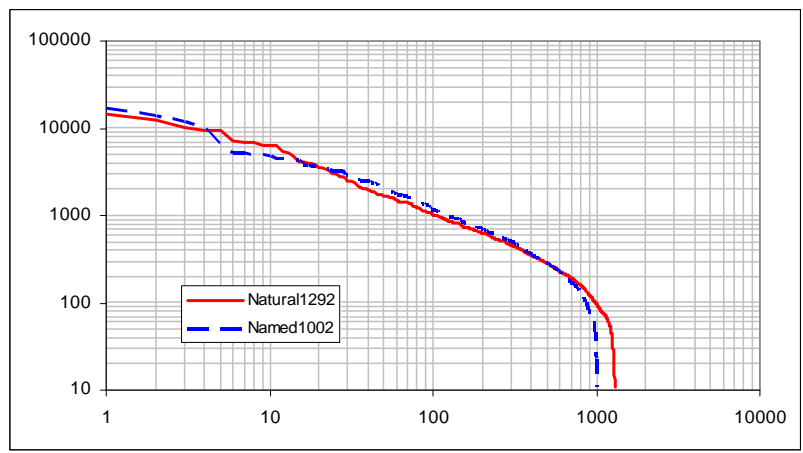

Figure 2: (color online) Rank size relations of street length

The street network can be topologically represented by a connectivity graph, consisting of vertices, representing individual streets, and edges if two streets are intersected. The connectivity graph forms the base for the following analysis. For the topology based on the notion of natural streets, the average connectivity (c.f. Appendix for details about topological measures) is 3.3, and the number of streets whose connectivity is less than the average is 76.5%. On the other hand, for the topology based on named streets, the average connectivity is 3.5, and the number of streets whose connectivity is less than the average is 79%. This pattern is clearly reflected in the log-log plot (Figure 3), where x and y axes represent connectivity and its cumulative probability, respectively. In the figure, the two indicated percentages are inversed ones with respect to the above 76.5% and 79%. The log-log curves are pretty close to straight lines, whose slope is around 2.0. This is not particularly surprising to us, for well connected streets tend to be lengthy, and connectivity and length are likely to be significantly correlated.

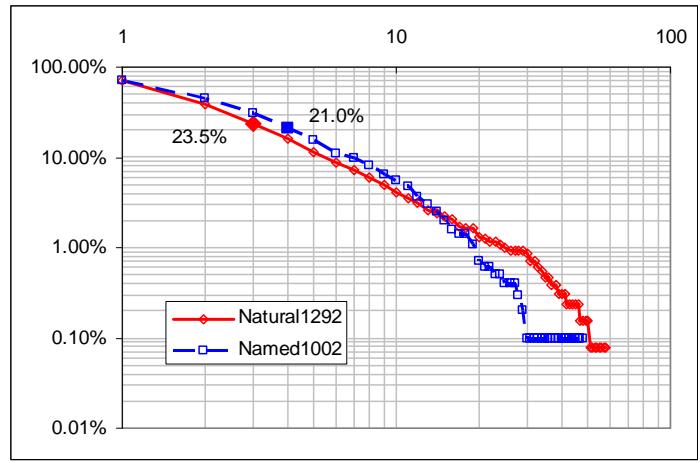

Figure 3: (color online) Log-log plot of the cumulative connectivity with the natural and named streets (slope around 2.0)

The previous study, based on a very big sample, illustrated the same pattern characterized by the 80/20 principle (Jiang 2007), i.e., 80 percent of streets are less connected (below the average), while 20 percent of streets are well connected (above the average); out of the 20 percent, there is 1 percent of streets that are extremely well connected. Now we can identify another 20 percent of streets that are extremely shorter. The streets less than 100 meters actually form another level of hierarchy, i.e., the bottom 20%. The bottom 20% is in fact included in the bottom 80%. In addition, we observe that the length for the top 20% of streets constitutes about 60% of total length of all the streets, and the length of the bottom 80% of streets constitutes about 40% of total length of all the streets. To further illustrate the street hierarchies, we map the individual streets using the same color legend used by Google Maps. As noted in Figure 4 and 5, the overall patterns are the same. The 1% streets are partially similar to yellowish streets (those visually prominent) displayed in Google Maps. This similarity provides primary evidence as to how structurally important ones match up functionally important ones. However, the two patterns are not identical for both natural and named streets. Essentially the processes of forming natural and named streets are different from one to another, so we should not expect the same result.

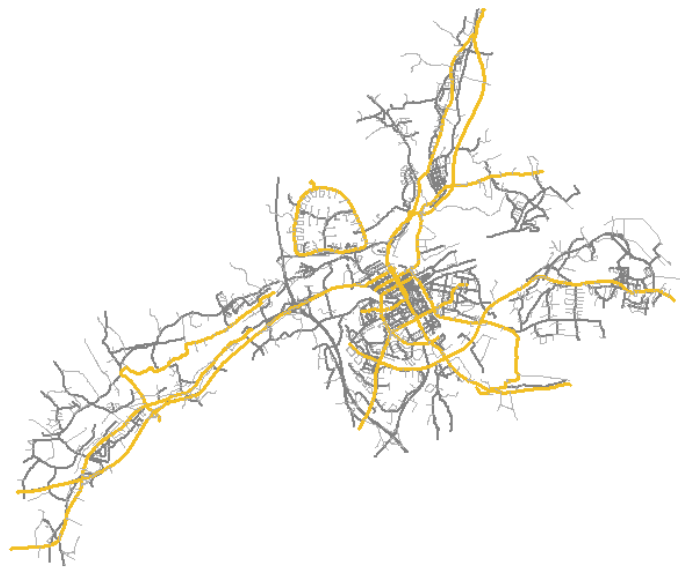

Figure 4: (color online) A spatial pattern of street hierarchies with the natural streets based on connectivity
(NOTE: yellow = top 1%, gray = top 20% exclusive of the top 1%, light gray = bottom 80%)

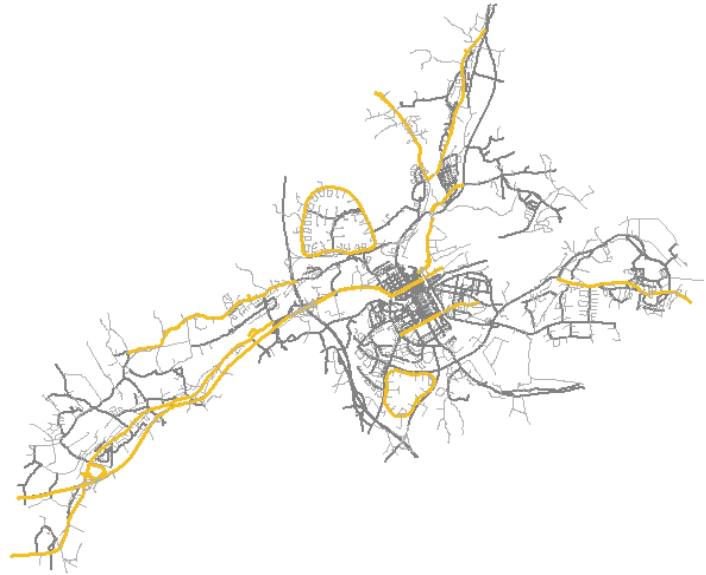

Figure 5: (color online) A spatial pattern of street hierarchies with the named streets based on connectivity
(NOTE: yellow = top 1%, gray = top 20% exclusive of the top 1%, light gray = bottom 80%)

Betweenness centrality (c.f. Appendix for more details) is used to identify those streets which have a bridging role between different topological shortest paths. The histogram of the distribution of betweenness is highly right-skewed, implying that while there are few streets of larger betweenness values, a significant majority of streets have smaller betweenness values. The log-log plot of the histogram follows quite closely a straight line except the falling part. The falling part represents those streets whose betweenness is less than 0.0015. Actually around 0.0015, all the streets can be partitioned again into the top 20% and bottom 80%. The 80% streets are with the falling part. Carefully checked with the street network, we noted that in terms of betweenness, the number of bottom 80% of streets is 925, while the number of bottom 80% of streets is 988 in terms of connectivity. In terms of both betweenness and connectivity, the number of bottom 80% of streets is 849. We can therefore note that both bottom 80% are significantly overlapped. In other words, those weak betweenness tend to be weak connected as well. Correlation analysis between connectivity and betweenness indeed illustrates the fact.

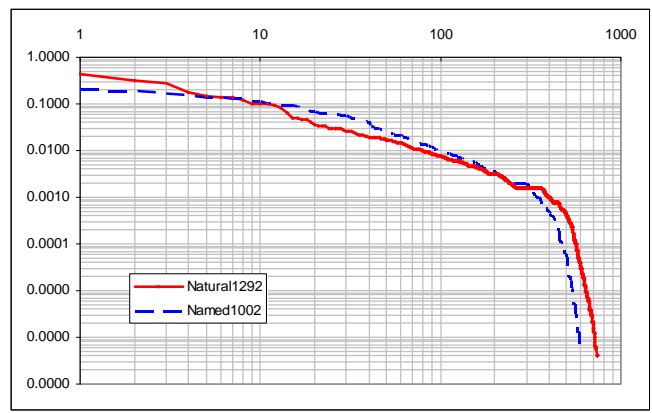

Figure 6: (color online) Rank size relations of betweenness with the natural and named streets
(NOTE: The falling part indicates the bottom 80% streets, while the relatively flat part is the top 20% streets. The partition for two parts is around betweenness centrality 0.0015.)

Table 1: Four non-exclusive levels of street hierarchies and their average statistics

| Levels | Natural streets | | | Named streets | | |
|---|---|---|---|---|---|---|
| | Length | Connectivity | Betweenness | Length | Connectivity | Betweenness |
| Top 1% | 7548.7 | 39.0 | 0.17 | 4977.7 | 24.4 | 0.105 |
| Top 20% | 1164.1 | 8.1 | 0.01 | 1439.8 | 9.0 | 0.02 |
| Bottom 80% | 230.4 | 1.8 | 0.00047 | 353.0 | 2.1 | 0.001 |
| Bottom 20% | 65.5 | 1.6 | 0.00034 | 67.8 | 1.7 | 0.0002 |

The above analysis proved that the Gävle streets have a coherent structure, for their geometric and topological properties follow scale-free property. There are many distinct scales with 4 orders of scale: from the shortest street of 10 meters to the longest street of 15 (natural street) or 17 (named street) kilometers with many more intermediate scales in between the shortest and longest, or alternatively in the same fashion, from the smallest connectivity of 1 to the most connected of 49 for the natural streets, and of 59 for the named streets. In the distinct scales, we identified four nonexclusive scales (Table 1), namely top 1%, top 20%, bottom 80% and bottom 20% (Note: the top 1% is included in the top 20%, and the bottom 20% is included in the bottom 80%). Our daily life starts with the bottoms, but without the tops our life would be chaotic. In the section that follows, we will demonstrate that the scaling law exists in reality with traffic flow, to prove how the functions of streets follow their morphology.

## 4. Traffic distributions among the individual streets

### 4.1. Data source and processing

The major data source was captured from taxi cabs through a collaboration project with a Gävle taxi company (TAXI Stor och Liten). All the cabs are equipped with GPS receivers. Considering the convenience of data transferring, we captured mobility information of cabs every ten seconds. The mobility information includes attributes of Date, Time, CarID, Status, Latitude, and Longitude (See table 2 for a sample). The mobility information projected in WGS84 can be easily georeferenced to relevant street networks for further analysis.

Table 2: Part of the mobility information for example
(NOTE: For the field status, 3 = Break, 4 = Not in service, 5 = Driving with taximeter on, 10 = Free)

| Date | Time | CarID | Status | Latitude | Longitude |
|---|---|---|---|---|---|
| 7/9/2007 | 9:12:13 AM | 7092 | 10 | 60.5450964114967 | 16.2862014808427 |
| 7/9/2007 | 9:12:13 AM | 7093 | 5 | 60.6625342510321 | 17.1193707029147 |
| 7/9/2007 | 9:12:13 AM | 7094 | 10 | 60.6143510482773 | 16.7774963417969 |
| 7/9/2007 | 9:12:13 AM | 7095 | 5 | 60.5303228042392 | 16.3005352058216 |
| 7/9/2007 | 9:12:23 AM | 7001 | 10 | 60.6644332550150 | 17.1526193658711 |
| 7/9/2007 | 9:12:23 AM | 7002 | 4 | 60.6485116622989 | 17.0140564481295 |
| 7/9/2007 | 9:12:23 AM | 7003 | 4 | 60.6620085380650 | 17.1522975007892 |
| 7/9/2007 | 9:12:23 AM | 7004 | 4 | 60.6699156902428 | 17.1868228952370 |
| 7/9/2007 | 9:12:23 AM | 7005 | 5 | 60.5724549434549 | 16.7568111458693 |
| 7/9/2007 | 9:12:23 AM | 7006 | 5 | 60.6539082668382 | 17.1307861844848 |

There are about 50 cabs in service daily, which covers two big towns (Gävle and Sandviken) and five small towns (Valbo, Kungsgården, Västerberg, Storvik, and Hofors) (Figure 7). In the current study, we concentrate on the city of Gävle, which is a middle city of 70 000 inhabitants.

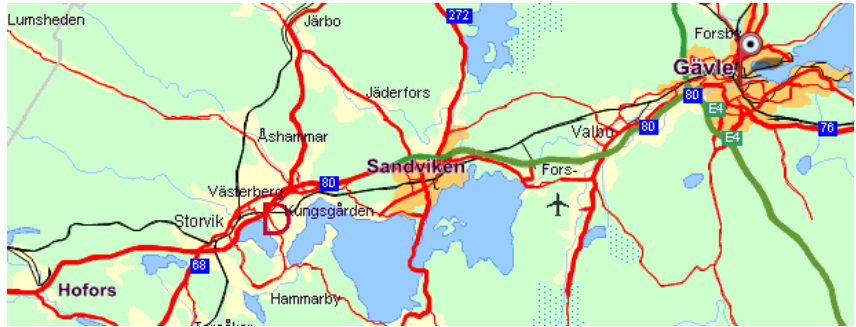

Figure 7: (Color online) Service coverage by TAXI Stor och Liten (the map is copyrighted by eniro.se)

We chose one week between the 1$^{st}$ and 7$^{th}$ of October 2007 for detailed analysis and observations. The raw data were processed to get rid of redundant information, to make sure all mobility information is valid. For example, we noticed that some cabs send signals from the same location to the central server for a very long time. Clearly it is redundant mobility information, which must be filtered out. In accounting for the use of streets, there is an impact from the speed of the cabs, i.e., the higher the cabs' speed, the less the number of recorded locations. By considering the average speed difference from one street to another, a little adjustment was done to get rid of the speed effect before the further analysis. In order to analyze the flow distribution, we created buffers of 10-meters (at each side of a central line street) for the individual streets, and obtained intensity of street use by counting the number of coordinates or locations recorded by GPS logs. Using the point-in-buffer operation, there are about 100K recorded locations every day. We plotted the 100K locations along the time line of 24 hours, and noticed two rush hours respectively for week days and weekend days, pretty consistent for each category. We skip more details about the rush-hours pattern due to business sensitivity.

**4.2 Emergence of hierarchy with traffic flow**
The cleaned data are further examined to assess how many locations or coordinates are with each individual street, a kind of intensity of street usage. Before a detailed comparison, we visually examined the traffic distribution among the individual streets. Figure 8 illustrates a sort of density map for both locations and trails for a part of the downtown of the city. We can observe that some streets are intensively used, while others are much less so. Besides, we can also notice two hotspots: one at the centre of the downtown and another at the central railway station.

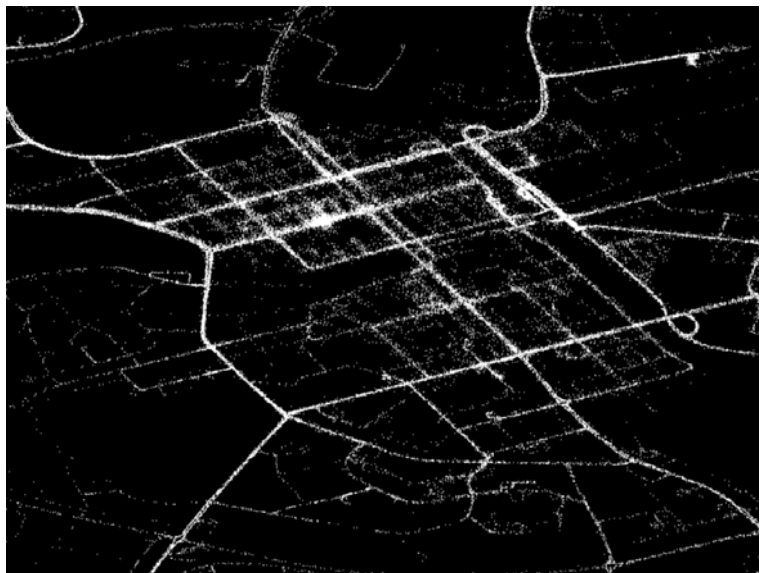
(a)

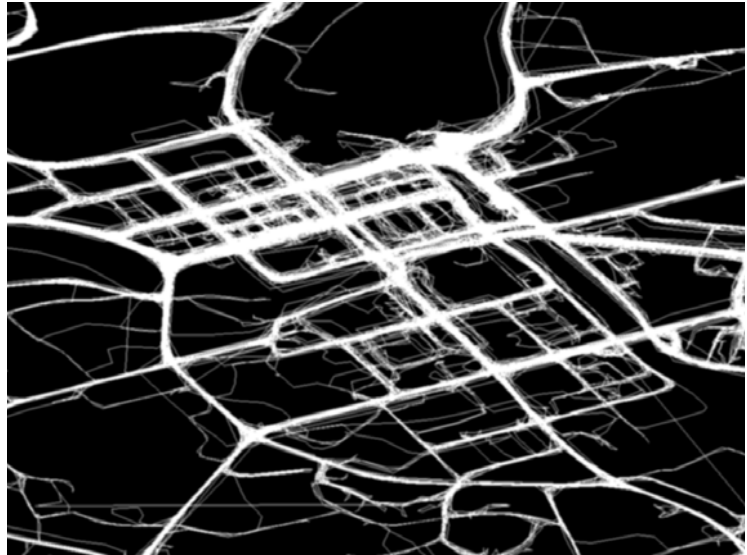

(b)

Figure 8: Street hierarchy shown in the downtown of the city with (a) coordinates map (b) and traces map
(NOTE: Some heavily traversed streets or locations indicated by white)

We further plot the number of recorded locations of the taxi cabs in individual streets. Clearly the flow distribution with the individual streets is strikingly skewed. The curves with the log-log plots are pretty consistent, very close to straight lines (figure 9). From the figure, we can observe that, for the natural streets, the curves appear to have two parts with different slopes (figure 9a). It illustrates the fact that all the natural streets can be put into two categories, each of which demonstrates regularity.

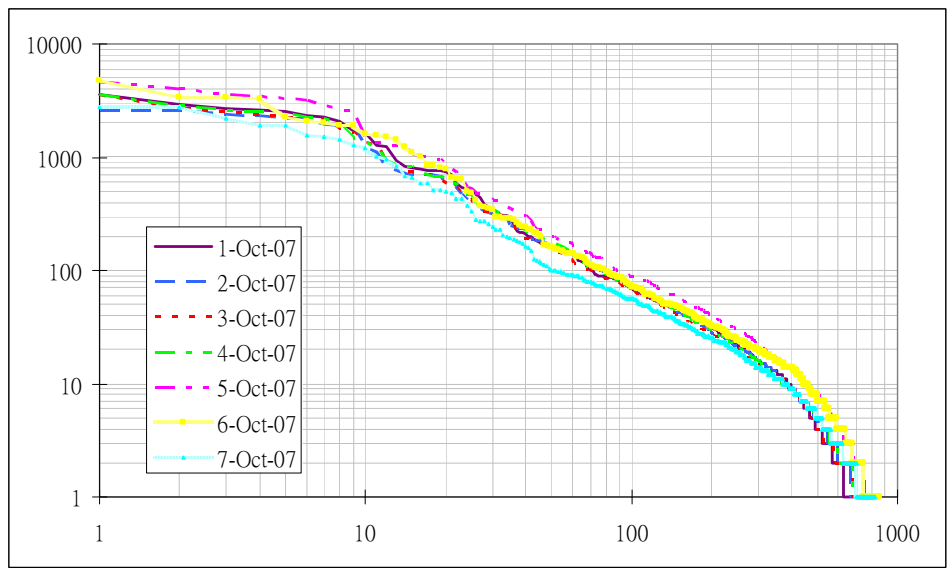

(a)

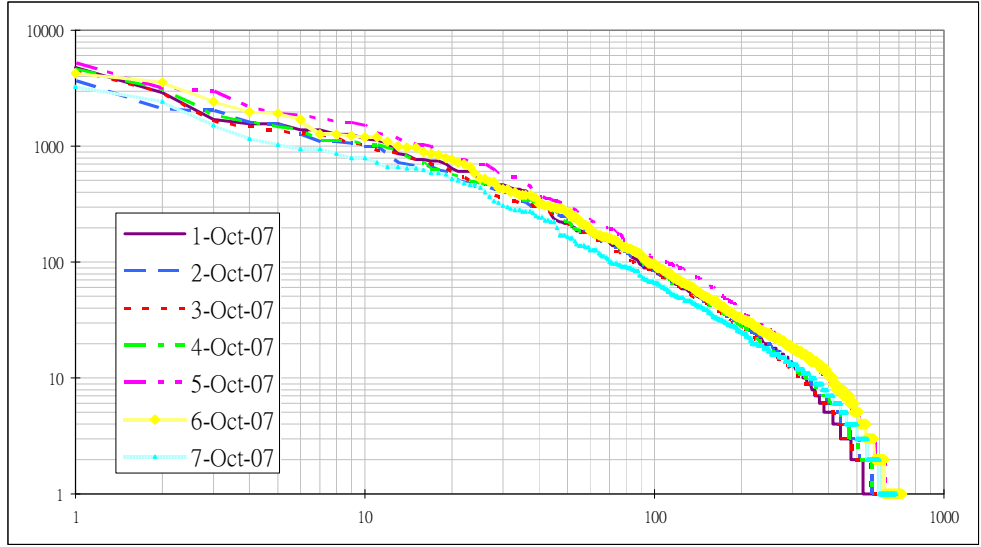

(b)

Figure 9: (color online) Rank-size relations of traffic flow among (a) the natural and (b) named streets

We further examined the flow distribution with reference to the four non-exclusive hierarchical levels. We found, in reference to Table 3, that the top 20% of streets indeed account for 80% of traffic, and the bottom 80% of streets account for about 15% and 20% of traffic, respectively, for natural and named streets. Furthermore, the top 1% of streets account for about 40% or 25% of traffic. This finding further proved our previous conjecture, i.e., a minority of streets account for a majority of traffic flow.

Table 3: The percentage of flow distribution among the streets from day to day

| | **Days** | **1-Oct** | **2-Oct** | **3-Oct** | **4-Oct** | **5-Oct** | **6-Oct** | **7-Oct** |
|---|---|---|---|---|---|---|---|---|
| **Natural streets** | **Top 1%** | 38 | 37 | 38 | 37 | 39 | 37 | 39 |
| | **Top 20%** | 85 | 85 | 85 | 85 | 86 | 86 | 86 |
| | **Bottom 80%** | 15 | 15 | 15 | 15 | 14 | 14 | 14 |
| | **Bottom 20%** | 2 | 2 | 2 | 2 | 2 | 2 | 2 |
| **Named streets** | **Top 1%** | 24 | 25 | 24 | 25 | 25 | 24 | 25 |
| | **Top 20%** | 82 | 81 | 81 | 81 | 82 | 82 | 81 |
| | **Bottom 80%** | 18 | 19 | 19 | 19 | 18 | 18 | 19 |
| | **Bottom 20%** | 2 | 2 | 2 | 2 | 2 | 2 | 2 |

Throughout the paper, we illustrated street hierarchies, an ordered hierarchical structure, from both geometry and topology perspectives. We further illustrated that traffic flow demonstrates a similar hierarchy. Our study proves again that how the function of the streets follow their morphology. At a much more detailed level, we examined how traffic flow is distributed among the four levels of hierarchy.

## 5. Discussions on the related work

The street hierarchies we found in the paper have far reaching implications to the understanding of the hierarchy of cities. In this connection, there has been much related work, mainly on theoretical conjectures, which deserves some further discussions. Based on the work of various researchers including Christopher Alexander (2004), Michael Batty (Batty and Longley 1994), and Bill Hillier (Hillier and Hanson 1984), Salingaros (2005) put forward a theory of the urban web from a network perspective. One of the key concepts of the theory is hierarchy, along with nodes and connections that form an urban web. The urban web self-organizes itself as a hierarchical structure at different levels of scale, from the smallest scales, and progressing up to the higher scales. Derived from biological and physical principles, Salingaros (2005) further suggested a multiplicity rule that can be applied to architectural and urban design. The multiplicity rule is a de facto power law we discovered for the street hierarchies. In this respect, our work can be considered to be a verification of the multiplicity rule from the perspective of urban streets. However, we tend to believe that the hierarchies of a city and its streets are self-organized from the bottom up through local interactions rather than the top down by design. A city indeed shows a hierarchical structure, and those city elements (in a broader sense rather than those suggested by Kevin Lynch) at a higher level of scale tend to form the image of the city (Lynch 1960) in our minds.

Not only constituents (e.g. streets or buildings) within a city, but also cities within a country or region are hierarchically organized, and they form intra-city and inter-city hierarchies, respectively. Cities within any country or all over the world are hierarchically organized, which means that there are many more small cities than large ones, following Zipf's law or Pareto's distribution. This universal law about city size distributions has been studied again and again (e.g., Zipf 1949, Simon 1955, Pumain 2006). A recent study developed a rather different approach to the city size distribution at a much more detailed level (Benguigui and Blumenfeld-Lieberthal 2007). The power law distribution of city size proved that a city system (composed of many cities in a country, region, or all over the world) is hierarchically organized, i.e., inter-city hierarchy. From the inter-city hierarchy to intra-city hierarchy, it reminds us of a recursive definition of a complex system, which reads as "a system that is composed of complex systems" (Batty 2006).

Hierarchies and communities have also featured the study of complex networks, in particular small world and scale free networks. Most of the real world networks exhibit scale-free properties in the sense that the connectivity of their nodes is very differentiated, demonstrating a hierarchical nature (Barabási and Albert 1999). There has been a vast amount of literature recently in modeling complex networks in detecting communities and hierarchical structures with the Internet and web. The metric clustering coefficient (c.f. Appendix for a definition) defined by Watts and Strogatz (1998) is found to be likely to have a power law distribution in most real world networks (Ravasz and Barabási 2003). However, this is not the case for street-street topology. As we experimented, the clustering coefficient does not follow the power law distribution as found in the topologies of many other complex networks. There are two opposite ways to explain it. First, street-street topology is indeed different from other network topologies that have emerged from biological, technological and social networks. Second, the metric does not capture the hierarchical nature of street-street topology. This issue certainly warrants further study in the future. Under the framework of complex networks, a series of studies about urban street networks (e.g. Kalapala et al. 2006, Porta, Crucitti and Latora 2006) have been carried out, and they have a close relationship with the paper.

**6. Conclusion**
We have in this paper studied the emergence of an ordered hierarchy with urban streets from the multiple perspectives of geometry and topology, as well as traffic flow in reality. We found throughout the paper that the distributions of some geometric and topological properties of urban streets are rather skewed, as well as that of traffic flow in individual streets. We identified four levels of scale, namely top 10%, top 20%, bottom 80% and bottom 20%, although they are not exclusive to each other. However, from the network topology (in terms of street-street intersection) as a whole, every level of scale is essential for human life. Our contact to the environment starts with the bottom 10% or bottom 80%, without them our life would be very difficult. In the same fashion, the top 1% or top 20% streets are for us to reach or get access to remote places, and without them, our life would be chaotic as well. This view has been articulated by Salingaros (2005) that living cities inevitably demonstrate an ordered hierarchical structure of connections at the different levels of scale. The street hierarchies illustrated reinforce the heterogeneity and diversity that characterize living cities, as argued by Jane Jacobs (1961) in her classic yet controversial work *Death and Life of Great American Cities*. The heterogeneity and diversity of streets become striking when we adopt a topological view from the perspective of individual streets.

We have illustrated the fact that how the function of streets follows up their topology and structure, i.e., vital streets tend to account for more traffic flow than those trivial ones. It is found to our surprise that the street hierarchies are a good indicator for traffic flow. That is, a majority of traffic occurs in the top 20% of streets, there is nearly no traffic in the bottom 20%, and interestingly, the top 1% of streets account for more than 20% of traffic. A minority of streets account for the majority of traffic flow sounds intuitive enough. However, we illustrated the fact in a rather precise fashion, i.e., 1 percent of top streets account for more than 20 percent of traffic, and 20 percent of top streets account for 80 percent of traffic. This finding is well supported by geographic information analysis, both from the structural analysis of urban street networks in terms of street-street topology, and empirical observation about traffic flow. A possible recommendation, derived from this finding, for planners and traffic engineers is that any redevelopment scenario must not violate the irony law of scaling. From another perspective, the scaling law provides a means to assess whether or not a certain scenario is acceptable for maintaining a healthy urban fabric.

**Acknowledgment**
This study was financially supported by research grants from the Hong Kong Polytechnic University, and the Swedish Research Council FORMAS. The GPS tracking dataset was kindly provided by the Gävle taxi company (TAXI Stor och Liten) in Sweden. In addition, the author would like to thank Junjun Yin, Sijian Zhao and Hong Zhang for their research assistance. Three referees provided some constructive comments. However, any shortcoming or inadequacy is the author's responsibility.

**Appendix: An introduction to topological measures used in the paper**

To complement the main text, this appendix aims to introduce the four topological measures examined or related to the paper. The first two are connectivity (or degree) and betweenness centrality. They were examined in a rather detailed fashion in the paper. The second two - clustering coefficient (CC) and closeness centrality, are just briefly mentioned or did not mention at all, because of their lack of a scaling property. In what follows, we try to avoid mathematics, and instead adopt a kite-shaped graph (Figure A-1, Krackhardt 1990) to illustrate the concepts and their differences (Table A-1).

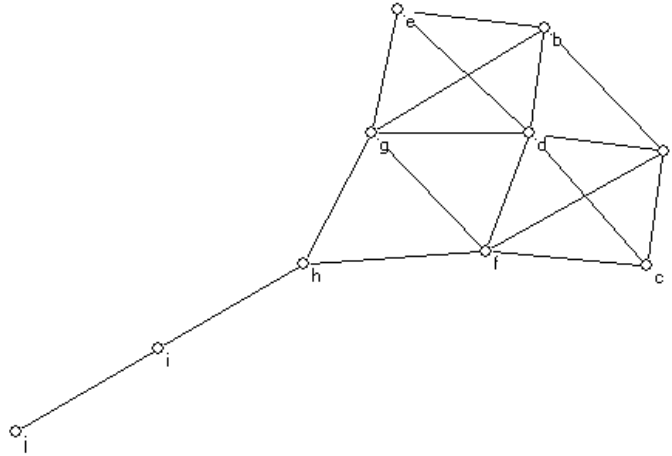

Figure A-1: The kite-shaped graph

Centrality measures were initially developed in social network analysis (Freeman 1979). They are used to describe the status of individual nodes within a graph from different perspectives. A degree centrality measures to what extent a node is connected directly by other nodes, thus a local measure in nature. The directed connected nodes are often called neighbors. The connectivity of individual streets used in this paper is in fact the degree of the nodes in the corresponding dual graph consisting of nodes representing streets, and links if two streets are intersected. A betweenness centrality measures to what extent a node is between two parts of a network. Those nodes that have a higher betweenness are those, without which a network would be separated into two parts. For example, the degree of node *h* is just 3, which implies it is not so important from a local perspective, but it is the most important one in the sense of betweenness. A closeness centrality measures to what extent a node is close to all other nodes, thus a global measure in nature. The nodes with highest closeness tend to be in the centre of a graph, e.g. nodes *f* and *g* with the kite graph.

Table A-1: Centralities and clustering coefficient for the nodes with Kite graph

| Node | Degree | Betweenness | Closeness | CC |
|---|---|---|---|---|
| a | 4 | 0.023 | 0.529 | 0.667 |
| b | 4 | 0.023 | 0.529 | 0.667 |
| c | 3 | 0.000 | 0.500 | 1.000 |
| d | 6 | 0.102 | 0.600 | 0.533 |
| e | 3 | 0.000 | 0.500 | 1.000 |
| f | 5 | 0.231 | 0.643 | 0.500 |
| g | 5 | 0.231 | 0.643 | 0.500 |
| h | 3 | 0.389 | 0.600 | 0.333 |
| i | 2 | 0.222 | 0.429 | 0.000 |
| j | 1 | 0.000 | 0.310 | 0.000 |

Clustering coefficient measures to what extent a node gets clustered, and it is defined by the probability that two neighbors of a given node are linked together. Taking nodes *c* and *e* for example, each has three neighbors, which are all directed linked each other, so the CC is 1. At another extreme, for nodes *i* and *j*, their neighbors if any are not directed linked at all, thus their CC are 0. The computation of the above measures with the study was done using Pajek (Nooy, Mrvar and Batagelj 2005).